# Unconditionally secure quantum key distribution over 50km of satndard telecom fibre

C. Gobby,* Z. L. Yuan and A. J. Shields

*Toshiba Research Europe Ltd, Cambridge Research Laboratory,*

*260 Cambridge Science Park, Milton Road, Cambridge, CB4 0WE, UK*

*Abstract:* We demonstrate a weak pulse quantum key distribution system using the BB84 protocol which is secure against all individual attacks, including photon number splitting. By carefully controlling the weak pulse intensity we demonstrate the maximum secure bit rate as a function of the fibre length. Unconditionally secure keys can be formed for standard telecom fibres exceeding 50 km in length.

*Introduction:* Quantum cryptography (QC) is often described as the first direct application of Quantum Mechanics. It provides a way of forming a secret key shared by two users (referred to as Alice and Bob) at either end of a communication channel [1]. The technique relies upon encoding the bit material for the key upon individual quanta, such as single photons. By encoding the information using non-orthogonal bases, the users can ensure that any attempt by a third party (Eve) to measure the encoded single photons, unavoidably alters their encoded state, resulting in errors in the shared key. Thus by monitoring the error rate in the formed key, Alice and Bob can detect any unauthorized intrusion on the communication channel.

In the real world there will also be errors due to imperfections in the system, such as detector noise and stray light [2]. These errors are indistinguishable from those caused by Eve. We therefore assume that the error rate determined for each key derives entirely from Eve. A classical protocol called privacy amplification [3] can be used to exclude any information potentially known by Eve, as implied by the measured error rate, and thereby guarantee the secrecy of the key. Ultimately, QC seeks to deliver 'unconditional secrecy', for which this guarantee is independent of any assumptions about the resources available to an eavesdropper.

The technology for performing QC across fibre optic cables is now relatively mature [2,4], with fibre lengths in excess of 120km reported [2]. The majority of demonstrations reported thus far have used weak laser pulses, rather than true single photon carriers. Even with strong attenuation, multi-photon pulses will be present in the signal pulses. These multi-photon pulses lead to a security loophole which Eve can exploit and theoretically employ a powerful eavesdropping attack known as photon number splitting (PNS) [7].

In this paper, we explore the practical limits for weak pulse QC using standard optical fibres. Keys, secure from any attack on individual bits (including the PNS attack) are formed using appropriate attenuation of the laser pulse and privacy amplification. We show that unconditionally secure key distribution is possible, and experimentally demonstrate the maximum secure bit rate as a function of the fibre length.

*The PNS attack:* The optimal eavesdropping attack on a weak pulse QC system is thought to be the PNS attack. Eve blocks all the single photon pulses and splits one photon from each multi-photon pulse and stores this for later measurement when Alice and Bob perform basis reconciliation. To hide her presence, Eve must ensure that Bob's detection rate is not altered, which she may do by replacing the channel to Bob with one of lower loss. Although the technology to perform such an attack does not exist at present, it is important to consider all theoretical forms of attack to guarantee the unconditional secrecy of the key.

The probability of Alice emitting a multi-photon pulse ($S_\mu$) is approximately $\mu^2/2$ for an ideal coherent source; where $\mu$ is the photon flux per clock cycle used by Alice. To safeguard against the PNS attack a simple criterion must be fulfilled: the rate of multi-photon pulses sent by Alice must be less than Bob's photon detection rate. Thus for each fibre length there is an optimal value for $\mu$ which provides the maximum bit rate secure against the PNS attack.

*Optimal photon flux calculation:* The probability that Bob detects a photon or an erroneous count, $P$, is given by

$$P = \mu\eta 10^{\left(-\frac{\alpha l}{10}\right)} + d \qquad (1)$$

where $\alpha$ is the fibre attenuation rate in dB/km and $\eta$ is Bob's detection efficiency, which combines the transmission loss of Bob's apparatus and the quantum efficiency of the detectors and $d$ is the erroneous count probability. Alice and Bob sift the measurement results to retain only the bits where they have used the same bases, resulting in a sifted bit rate of ½ $P$ for the BB84 protocol and a quantum bit error rate (QBER) of $e=$½ $d/P$. The QBER is defined as the ratio of wrong bits to the total number of bits formed in the sifted key. They then carry out error correction [6] to remove the discordant bits and privacy amplification [3] to exclude any information potentially known to Eve. The secure bit rate per clock cycle is then given by [7]

$$G = \frac{1}{2}P\left\{\frac{P-S_\mu}{P}\left(1 - Log_2\left[1 + 4e\frac{P}{P-S_\mu} - 4\left(e\frac{P}{P-S_\mu}\right)^2\right]\right) + f(e)[e\log_2 e + (1-e)\log_2(1-e)]\right\}$$

(2)

The first term in Eq.2 describes the reduction in the bit rate due to privacy amplification, where it is assumed that Eve can exploit all the multi-photon pulses to determine some of the bits, as well as some other form of individual attack which leads to the QBER, $e$. Notice that for the secure bit rate to remain positive, $P > S_\mu$, reiterating the criterion above. The second term in Eq.2 describes the reduction in bit rate due to correcting the errors in the raw sifted key. The efficiency of the error correction depends on the particular algorithm used. The Shannon limit provides $f(e)=1$, however, all known algorithms are less efficient. The well known Cascade protocol, gives f[e] = 1.16 for e < 5 % [6].

Figure 1 shows the calculated dependence of the secure bit rate upon the fibre length and photon flux per clock cycle used. The calculation takes the parameters pertinent to the experiment described later. For each fibre length up to 56 km, there exists a range of $\mu$ for which unconditionally secure key distribution is possible. The upper bound on $\mu$ is a consequence of the PNS attack on the multi-photon pulses, while the lower bound on $\mu$ derives from an increase in the QBER due to the erroneous counts in the detectors. Between these two limits, there exists an optimal laser intensity, for which the maximal secure bit rate can be obtained. The dashed line in Fig. 1 shows the optimal value of $\mu$ as a function of fibre length. The optimal $\mu = 0.046$ photons/clock cycle at 1 km, declining to 0.0042 photons/clock cycle at 50 km.

Notice that the values of $\mu$ used in previous demonstrations [2,4] of lie outside the secure bounds in Fig.1. The keys formed in these experiments are vulnerable against the PNS attack, thus highlighting the vulnerability of weak pulse QC.

*Experimental set-up:* Our system is based upon a time/polarisation division Mach-Zender interferometer [2]. Signal pulses are generated by a 1.55 µm DFB laser diode operating at 2 MHz with a pulse width of 80 ps. The pulses are attenuated to the optimal level required for the maximum unconditionally secure bit rate, as indicated by the dashed line in Fig. 1. The weak coherent pulses are then multiplexed with strong pulses from a 1.3 µm clock laser, which serves as a timing reference. Phase modulators controlled by custom electronics in the two interfering routes are used to encode the bit information using BB84 protocol [1].

The signal photons are detected by InGaAs avalanche photodiodes, cooled to an approximate temperature of -100ºC, operating in gated mode with a gate width of 3.5 ns and an excess voltage of 2.5 V. Our detectors typically have a dark count probability of $10^{-7}$ per ns, along with a detection efficiency of ~12% at 1.55 µm. The total erroneous count probability is ~8 x $10^{-7}$ per gate, which is due to both detector dark counts and stray light from the 1.3 µm clock laser. Bob's overall detection efficiency is 4.5%, which includes both the efficiency of the detector and optical losses in Bob's apparatus.

*Results and discussion:* The measured QBER, as shown in the inset of Fig. 2, remains virtually constant up to 30 km. The fibre attenuation is not so severe and therefore a reasonably strong $\mu$ can be used. As a result, the photon detection rate remains significantly higher than the erroneous count rate. The main contribution to the measured QBER is believed to derive from inaccuracy of modulator voltages and interferometer phase drift. Over 30 km the measured QBER starts to increase with increasing fibre length, because the detector erroneous counts are no longer negligible compared to the signal photon count rate, which is reduced by fibre attenuation and the higher attenuation required for unconditional security at these lengths. A calculation of the QBER, which includes the fibre attenuation and the measured detector erroneous rate, shown as the solid line in the inset of Fig. 2, agrees well with the experimental data.

Figure 2 shows the sifted bit rate (solid symbols) as a function of the fibre length. The sifted bit rate falls with increasing fibre length at a rate of ~0.5dB/km, which is much higher than the fibre attenuation of ~0.2dB/km. This is due to the extra attenuation of the weak pulses introduced to defeat the PNS attack., indicating the cost of unconditional security The sifted bit rate is 700 bits/s at 4.4km link, but falls to 20 bits/second for a 50-km link. Notice the close agreement between the experimental points and the calculated data (shown as a solid line).

Error correction [6] and privacy amplification [3] were applied to the sifted bits to form an unconditionally secure key. The resulting secure bit rates are shown in Fig. 2. The secure bit rate is up to 300 bits/s for short fibre lengths, decreasing to 2.1 bit/s for 44 km. The longest fibre length for which we could form an unconditionally secure key is 50.6km. A good agreement is found between the experimentally measured secure bit rates and a calculation based upon Eq. 2 shown as the dashed line.

*Summary:* In summary, we have demonstrated a weak pulse QC system using the BB84 protocol that is secure from the PNS attack. An analysis of the secure bit rate shows that there is a range of weak pulse intensities allowing a secure key to be formed. Using the optimal weak pulse intensity, we have demonstrated the maximum secure bit rate as a function of fibre length. The secure bit rates achieved after error correction and privacy amplification agree closely with the calculated values. The weak pulse system can operate with fibres up to 50.6 km in length.

## REFERENCES

[*] also at Cavendish Laboratory, University of Cambridge, Cambridge CB3 0HE, UK

1. Bennett, C.H., and Brassard, G.: 'Quantum cryptography', Proc. IEEE Int. Conf. on Computers, Systems and Signal Processing, Bangalore, India, 1984, pp.175-179

2. Gobby, C., *et al.*: 'Quantum key distribution over 122 km standard telecom fiber', *Appl. Phys. Lett.*, 2004, **84**, pp. 3762-3764

3. Bennett, C.H., *et al.*: 'Generalized privacy amplification', *IEEE Trans. Inform. Theory*, 1995, **41**, pp. 1915-1923

4. Stucki, D., *et al.*:'Quantum key distribution over 67 km with a plug & play system', *New J. Phys.*, 2002, **4**, pp. 41.1-41.8

5. Brassard, G., *et al.*: 'Limitations on practical quantum cryptography', *Phys. Rev. Lett.*, 2000, **85**,1330-1333

6. Brassard, G., and Savail, L.: 'Secret-key reconciliation by public discussion', *Lect. Notes Comp. Sci.*, 1994, **765**, pp. 410-423

7. Lütkenhaus, N.: 'Security against individual attacks for realistic quantum key distribution', *Phys. Rev. A*, 2000, **61**, p. 052304

*Figure captions:*

**Figure 1** A contour plot of the secure bit rate as a function of the average number of photons per clock cycle ($\mu$) and the fibre length. In the calculation, $\eta$=0.045, $\alpha$=0.21dB/km, $f(e)$=1.18 and the modulation error = 3.0%.

**Figure 2** Plot of the measured (solid symbols) and calculated (solid line) sifted bit formation rate as a function of fibre length. The unconditionally secure net bit rate (open symbols) shows good agreement with a calculation based on Eq.2 (dashed line). The inset shows the measured (symbols) and calculated (line) QBER as a function of fibre length.

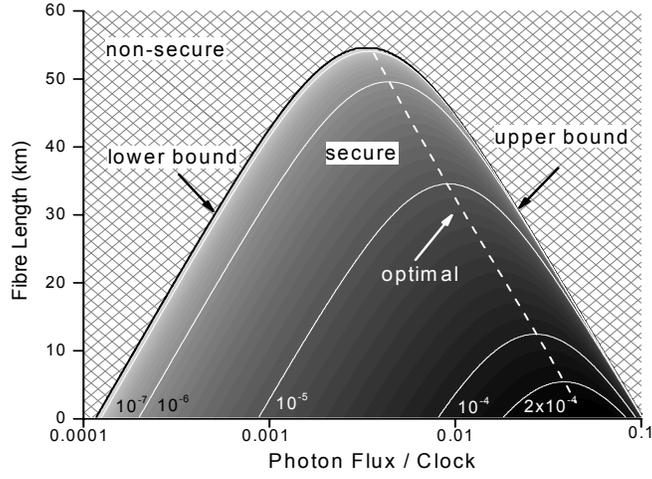

Figure 1 Gobby et al

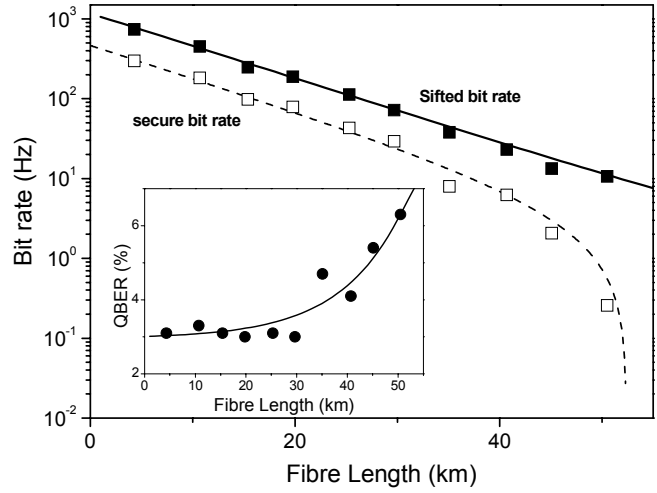

Figure 2 Gobby et al

9